\documentclass[reprint,amsmath,amssymb,aps]{revtex4-2}

\usepackage[pdftex]{graphicx}
\usepackage{dcolumn}
\usepackage{bm}
\usepackage{amsmath}
\usepackage[usenames,dvipsnames]{xcolor}
\usepackage{mathtools}
\usepackage[colorlinks=true,citecolor=blue,linkcolor=blue]{hyperref}
\usepackage{bbm}
\usepackage{comment}
\usepackage{upgreek}
\usepackage{esint}
\usepackage{soul}
\usepackage[normalem]{ ulem } 
\usepackage{floatrow}
\usepackage{verbatim}
\usepackage{braket}
\usepackage{amsmath, amssymb}
\usepackage{array}
\usepackage{algorithm}
\usepackage{algpseudocode}
\usepackage{graphicx}
\usepackage{xcolor} % put this in your preamble
\usepackage{enumitem}
\usepackage{subcaption}
\usepackage{multirow}
\usepackage{booktabs}
\usepackage{array} % preamble

\usepackage[colorlinks=true,citecolor=blue,linkcolor=blue]{hyperref}
\newcolumntype{M}[1]{>{\centering\arraybackslash}m{#1}}
\newcolumntype{L}[1]{>{\raggedright\arraybackslash}m{#1}}
\newcolumntype{R}[1]{>{\raggedleft\arraybackslash}m{#1}}

\setlist[itemize]{parsep=1em, itemsep=1em} % Adjust spacing for itemize
\setlist[enumerate]{parsep=0em, itemsep=1em} % Adjust spacing for enumerate

\newcommand{\PreserveBackslash}[1]{\let\temp=\\#1\let\\=\temp}
\newcolumntype{C}[1]{>{\PreserveBackslash\centering}p{#1}}

\newcommand{\rom}[1]{\uppercase\expandafter{\romannumeral #1\relax}}
\newcommand{\umq}{U_\text{MQ}}
\newcommand{\fanout}{\textsf{\textsc{fanout}}}

\begin{document}

\title{Phase gadget compilation of quantum circuits using multiqubit gates}

\author{Jonathan Nemirovsky}
\author{Maya Chuchem}
\author{Lee Peleg}
\author{Yakov Solomons}
\author{Amit Ben Kish}
\author{Yotam Shapira}\email{yotam.shapira@quantum-art.tech}

\affiliation{Quantum Art, Ness Ziona 7403682, Israel}

\begin{abstract}
	Quantum circuit synthesis and compilation are critical components in the quantum computing stack, both for contemporary quantum systems, where efficient use of limited resources is essential, as well as for large-scale fault-tolerant platforms, where computation time can be minimized. The specific characteristics of the quantum hardware determine which circuit designs and optimizations are feasible. We present a phase-gadget based method for compilation of quantum circuits using programmable multiqubit entangling gates, that are native, among others, to trapped-ions quantum computers. We use phase-gadgets in order to generically reduce circuit depths and efficiently implement them with few, high-fidelity, multiqubit gates. We test our methods on a large set of benchmark circuits and demonstrate generic circuit depth reduction and implementation error reduction.
	
\end{abstract}

\maketitle

\section{Introduction}
Quantum computers have the potential to deliver meaningful improvements to computational tasks across a broad range of scientific and technological domains. Some advantages may already be accessible with contemporary hardware \cite{zimboras2025myths,shaydulin2024evidence}, but realizing them requires careful alignment between hardware capabilities and algorithmic design. Consequently, quantum circuit synthesis and compilation constitute crucial components of the quantum computing stack. This is as well true in the fault-tolerant regime, where the overhead of quantum error correction is substantial and optimizing circuit runtimes is essential.

Clearly, the properties of the physical platform on which the quantum computation is realized determine what sort of circuit synthesis and optimizations can be performed. For example, in qubits encoded by superconducting circuits, entanglement operations are typically local \cite{tan2020optimal}, however can be performed in parallel. Alternatively, in the quantum charge coupled device, arbitrary qubit pairs can be entangled by shuttling them to a dedicated `entanglement region', however the number of entanglement gates that can be performed in parallel is limited by the number of these entanglement regions \cite{moses2023race}.

Here we focus on programmable multiqubit entanglement gates that enable arbitrary coupling between all $\binom{N}{2}=N\left(N-1\right)/2$ pairs of a $N$-qubit quantum register. Specifically, we consider the gate,
\begin{equation}
	\umq\left(\varphi\right)=\exp\left(i\sum_{n,m=1}^N \varphi_{nm} Z_n Z_m\right),
	\label{eqUmq}
\end{equation}
where $Z_n$ is the $Z$-Pauli operator, acting on the $n$'th qubit and $\varphi\in\mathbb{R}^{N\times N}$ is an arbitrary user-determined matrix, such that $\varphi_{nm}$ is a $Z\otimes Z$ entanglement phase between the $n$'th and $m$'th qubits.

Gates of the form of Eq.~\eqref{eqUmq} arise naturally in long crystals of trapped-ions \cite{feng2023continuous,pogorelov2021compact,yao2022experimental,guo2024site,schwerdt2024scalable}, where the arbitrary coupling is due to the underlying long-range Coulomb coupling between the ions. Indeed, the realization of such gates is intensively researched, both theoretically \cite{grzesiak2020efficient,shapira2020theory,shapira2023fast,lu2019global,bassler2023synthesis,solomons2025programmablequantumcomputingtrappedions} and experimentally \cite{grzesiak2020efficient,lu2019global,lu2025implementing,shapira2025programmable}. We remark that additional quantum computing modalities have the potential to generate such programmable multiqubit interactions \cite{evered2023high,jeremy2021asymmetric,cooper2024graph,yongxin2025constant}, and can as well benefit from our method.

Clearly, when concatenated with arbitrary single-qubit gates, the variety of gates enabled by $\umq$ forms an immensely over-complete universal gate-set, leaving much room for optimization of the target quantum circuits. Specifically, these gates hold the potential for a quadratic (in $N$) speedup in circuit depth \cite{yin2025fastquantumcomputationalltoall}. Indeed, several speedups enabled by gates of the form of $\umq$ are already known, e.g. compared to platforms that only enable serial gates. Specifically multiply-controlled Toffoli (Toffoli-$n$) is done in logarithmic depth (compared to linear depth) \cite{bravyi2022constant}, quantum Fourier transform can be done in constant or linear depth (compared to quadratic depth) \cite{bassler2023synthesis,hoyer2003quantum,hoyer2005quantum},  Clifford circuits, including stabilizer circuits for QEC, in constant depth (compared to linear depth) \cite{bravyi2022constant,cleve2025improvedcliffordoperationsconstant,nemirovsky2025reduced}, Quantum Volume circuits in linear depth (compared to quadratic depth) \cite{nemirovsky2025efficient} and random unitaries in constant depth (compared to logarithmic depth) \cite{foxman2025random}. Furthermore, generic speed-ups in quantum optimization and compilation are expected \cite{galicia2020enhanced,chandarana2025runtimequantumadvantagedigital,bassler2023synthesis,villoria2025optimizationsynthesisquantumcircuits}.
	
Here we present a compilation method for generic quantum circuits that makes use of multiqubit gates, $\umq$ as in Eq.~\eqref{eqUmq}, and arbitrary single-qubit rotations. Our method acts to reduce the overall multiqubit gate count. Furthermore, our method also reduces the total drive power used to generate multiqubit gates, which in turn reduces gate errors and increases overall performance. We further show that with the addition of a single auxiliary (ancilla) qubit all of the multiqubit gates realize Clifford gates, which are substantially easier to benchmark and optimize \cite{huang2022foundations}, and are advantageous for implementations on logical qubits. We test our method on a set of benchmark circuits \cite{li2023qasmbench} and demonstrate a generic reduction of circuit depth by a factor of $\sim15$, accompanied by a reduction of the relative circuit implementation errors by $\sim40\%$.

In general, our compilation operates by transforming input quantum circuits into a circuit composed of three layers. The first and last layer are both classical, i.e. they can be efficiently performed on a classical computer, and do not incur a quantum compute cost. The middle layer is composed of a series of phase-gadgets (PGs), i.e. exponentiation of Pauli strings with some rotation angles \cite{vandewetering2020zx,cowtan2020phase}. Specifically, we make use of $Z$-PGs, i.e. operators of the form 
\begin{equation}
	G_Z\left(\alpha,\{j_1,j_2,j_3,...\}\right)=\exp\left(i\alpha\frac{\pi}{2} Z_{j_1}Z_{j_2}Z_{j_3}\cdots\right),\label{eqGz}
\end{equation}
where $\alpha$ is the rotation angle and $J=\{j_1,j_2,j_3,...\}$ is a set of target qubit indices. We also make use of analogously defined $X$-PGs, $G_X$. Clearly such a decomposition is universal.

This decomposition is useful since $G_{X/Z}$ have a constant-cost implementation using $\umq$ (or equivalently using $\fanout$ gates). Specifically, any $G_{X/Z}$, regardless of $\alpha$ or $J$, can be implemented with two applications of $\umq$. Furthermore, provided a single auxiliary (ancilla) qubit a series of $M$ such PGs are implemented with only $M+1$ application of $\umq$. To utilize the large connectivity afforded by $\umq$ we commute many two-qubit CNOT operations to the first and last layer at the `price' of generating PGs with many connected qubits, i.e. large sets of qubit indices, $J$s, however since this does not affect the multiqubit gate count it is beneficial and improves the overall performance.

This compilation approach offers multiple degrees of freedom, e.g. each CNOT gate can be commuted to the beginning or to the of end of the circuit. In addition local blocks of $SU(4)$ operations can be decomposed in various ways, which affect the results. Thus our method includes polynomial-time optimization passes, which operate globally on the entire circuit.

We remark that PGs are widely used in research, especially using ZX-calculus for circuit simplification. Here, we do not use these approaches \cite{vandewetering2020zx,cowtan2020phase,wetering2025optimal,riu2025reinforcement,chen2025quantum,villoria2025optimization}, however they can be combined with our approach in a straightforward manner.

The remainder of the paper is ordered as follows: Section \ref{sec:Implementation} describes implementation of PGs with $\umq$, section \ref{sec:Comm} provide useful PG commutation rules, sections \ref{sec:SU4}, \ref{sec:CompPrim} and \ref{sec:CompGlob} describe the set of compilation primitives we use, section \ref{sec:Noise} discusses noise models and finally, section \ref{sec:Bench} provides benchmark data that demonstrates the efficacy of our method.

\section{Implementation of phase gadgets}\label{sec:Implementation}
A useful decomposition of PGs is given in terms of generalized CNOT gates, $C_{P_j\land Q_k}=\exp\left[i (I-P_j)(I-Q_k)\pi/4\right]$, with $P_j,Q_k\in\{X,Y,Z\}$ and $I$ the identity operator, such that the canonical CNOT gate with the $j$th qubit controlling the $k$th qubit is given by $C_{Z_j\land X_k}$. With this definition we write phase gadgets as (see the SM \cite{SM}),
\begin{align}
	G_P & \left(\alpha, J\right)= \nonumber \\
	 & \left(\prod_{j\in J\setminus{j^\ast}}C_{P_j\land Q_{j^\ast}}\right)G_P(\alpha,\{j^\ast\})\left(\prod_{k\in J\setminus{j^\ast}}C_{P_k\land Q_{j^\ast}}\right),\label{eqPGdecomp}
\end{align}

\begin{figure}[h]
	\centering
	\begin{subfigure}[a]{1.0\textwidth}
		\centering
		\includegraphics[width=\textwidth]{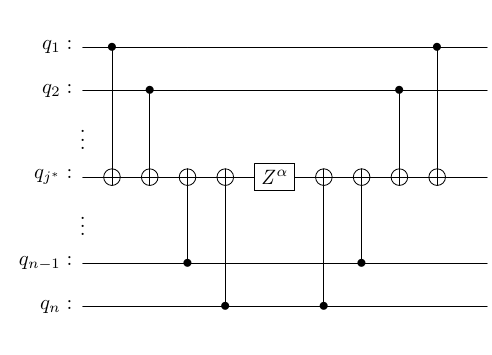}
		\caption{Implementation of $ G_Z(\alpha,\{q_1,...,q_n\}) $ phase gadgets.}
		\label{fig:sub1}
	\end{subfigure}
	\hfill

	\begin{subfigure}[b]{1.0\textwidth}
		\centering
		\includegraphics[width=\textwidth]{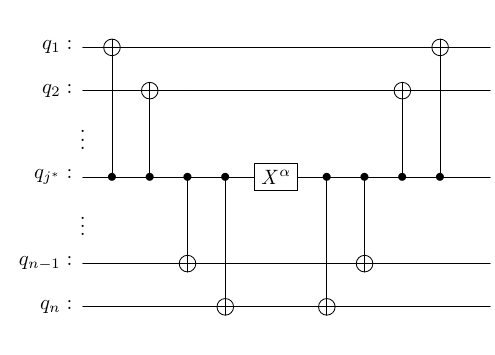}
		\caption{Implementation of $ G_X(\alpha,\{q_1,...,q_n\}) $ phase gadgets.}
		\label{fig:sub2}
	\end{subfigure}
		\caption{Implementation of phase gadgets using two multiqubit gates, $\umq$.}
		\label{fig:PG_with_CNOTS}
	\label{fig:main}
\end{figure}

where $j^\ast\in J$ is chosen arbitrarily and $Q$ as well is chosen arbitrarily such that $P\neq Q\in\{X,Y,Z\}$. We note that both products over CNOT gates in Eq.~\eqref{eqPGdecomp} form $\fanout$ operations, which are each implemented with a single application of $\umq$, and the middle phase-gadget, $G_P(\alpha,\{j^\ast\})=\exp\left(\frac{i\alpha\pi}{2} P\right)$, is in fact a simple single-qubit operation. Figure~\ref{fig:PG_with_CNOTS} shows realizations of $Z$ (a) and $X$ (b) phase-gadgets using this identity. With this construction, a series of $M$ alternating $X$ and $Z$ PGs are implemented with $2M$ applications of $\umq$.

\begin{figure}[h]
	\centering
	\begin{subfigure}[a]{0.66\textwidth}
		\centering
		\includegraphics[width=\textwidth]{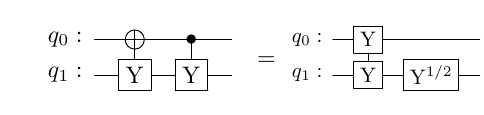}
		%\caption{\tiny {Implementation of $ \prod_{j\in J}C_{Z_j\land Y_a} $ with a single MQ layer.}}
		\label{fig:sub1}
	\end{subfigure}
	\begin{subfigure}[b]{1.0\textwidth}
		\centering
		\includegraphics[width=\textwidth]{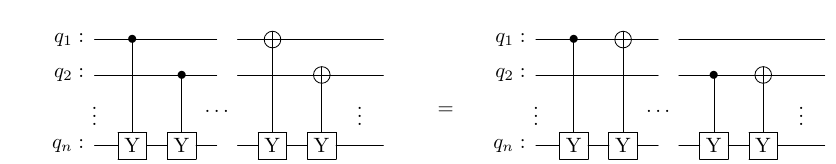}
		%\caption{\tiny {Implementation of $ \prod_{j\in J}C_{X_j\land Y_a} $ with a single MQ layer.}}
		\label{fig:sub1}
	\end{subfigure}
	\hfill
	% Second subfigure
	\begin{subfigure}[c]{1.0\textwidth}
		\centering
		\includegraphics[width=\textwidth]{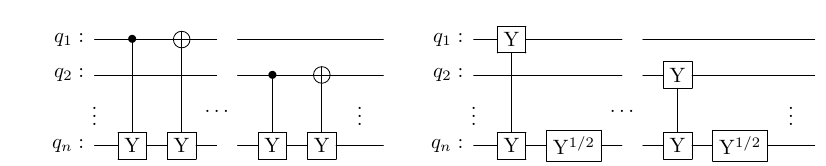}
		%\caption{\tiny {Implementation of $ \prod_{j\in J}C_{Z_j\land Y_a}  \prod_{k\in J}C_{X_k\land Y_a} $ with a single MQ layer.}}
		\label{fig:sub2}
	\end{subfigure}
	\caption{Sketch of the proof of the PG merge identity in Eq.~\eqref{eqPGmerge}. The basic identity used is that the circuit $C_{Z_j\land Y_k}C_{X_j\land Y_k}$ can be implemented with a single two-qubit $Y\otimes Y$ gate (top), equivalent to a $Z\otimes Z$ gate. When consecutive $G_X$ and $G_Z$ PGs appear, the CNOT layer at their interface consists of multiple such instances (middle), this can be rearranged, and cast a single multiqubit gate (bottom).}
	\label{fig:MERGE_CNOTS_NEW}
\end{figure}

An equivalent decomposition is achieved by considering an additional single auxiliary (ancilla) qubit, denoted by the index $a$. We implement the PGs using the same form of Eq.~\eqref{eqPGdecomp} but setting $j^\ast\mapsto a\notin J$, with the single qubit rotation $\exp\left(i\alpha\frac{\pi}{2}Z_a\right)$ operating on this auxiliary and setting the initial state of this qubit to $ \left|0\right>$ - i.e., the ground state of the Pauli-$Z_a$ operator. With this, the collective phase generated by the phase-gadget still operates on the system and qubit $a$ remains disentangled from the remaining qubits. This approach enables further compression of the multiqubit gate count in implementing a series of alternating $X$ and $Z$ PGs. Specifically, we set $Q_a=Y_a$ such that, e.g. for $G_Z(\alpha,J)$ followed by $G_X(\beta,K)$ the concatenation of CNOTs at the interface between the PGs yields,
\begin{equation}
	\prod_{j\in J}C_{Z_j\land Y_a}\prod_{k\in K}C_{X_k\land Y_a}\cong \umq\left(\varphi^{\left(J\cup K,\{a\}\right)}\right)\label{eqPGmerge},
\end{equation}
with the equivalence relation, `$\cong$', signifying equality up to single qubit gates (see the SM \cite{SM}) and the non-zero entries of the symmetric matrix $\varphi^{\left(J\cup K,\{a\}\right)}$ given by,
\begin{equation}
	\varphi^{\left(J\cup K,\{a\}\right)}_{n,m}=\begin{cases}
		\pi/4 & \text{if } n\in J\cup K \text{and } m=a, \\
		0  & \text{otherwise}.
	\end{cases},\label{eqPGmergeDetails}
\end{equation}
Clearly, the right-hand side of Eq.~\eqref{eqPGmerge} is implemented with a single application of $\umq$. With this, a series of $M$ alternating $X$ and $Z$ PGs are implemented with $M+1$ applications of $\umq$. Furthermore, since the phases in Eq.~\eqref{eqPGmergeDetails} are either 0 or $\pi/4$ then these multiqubit gates are Clifford operations, which are experimentally easier to calibrate and benchmark \cite{huang2022foundations}. In addition, the only non-Clifford operations required in these decompositions are arbitrary single-qubit rotations of the single auxiliary qubit. Figure~\ref{fig:MERGE_CNOTS_NEW} shows a sketch of the proof of Eq.~\eqref{eqPGmerge}, with a full derivation appearing in the SM \cite{SM}. 

In trapped-ions quantum computers the physical realization of $\umq\left(\varphi\right)$ is performed by applying electromagnetic fields that drive the qubits appropriately, generating the desired entanglement gate. While determining the specific details of the qubit drive involves solving a challenging optimal-control problem, the magnitude of the drive can be estimated directly from $\varphi$ \cite{shapira2023fast}. Namely, the drive power scales as the nuclear-norm of $\varphi$, i.e. the $L_1$ norm of its eigenvalues. Minimizing the total drive power used to implement the circuit is an additional performance merit of the circuit, as many sources of error and noise scale with the drive power, e.g. depolarization due to photon scattering.

Using this power estimate, the implementation of the `all-to-one' gates in Eq.~\eqref{eqPGmerge} requires a drive power that scales as $\sqrt{|J\cup K|}$, i.e. the square-root of the number of qubits in the union of $J$ and $K$. Considering the same scaling, we note that an alternative implementation of the same gate with sequential two-qubit gates will require a total power that scales as $|J\cup K|$, demonstrating an additional advantage of $\umq$ in terms of drive power.

\section{Phase gadget commutation rules}\label{sec:Comm}
Our compilation method commutes CNOT gates to the edges of the circuit, realizing the three layers discussed above. We show that in the phase-gadget framework, commutation of CNOT gates acts to generate, modify or annihilate PGs. A summary of these commutation rules is provided below, with a more detailed account provided in Refs.~\cite{SM,cowtan2020phase,bravyi2021hadamard}.

A canonical CNOT, $C_{j,k}=C_{Z_j\land X_k}$, commutes with $Z$ PGs as,
\begin{equation}
	\begin{cases}
		C_{j,k}G_Z(\alpha,K)=G_Z(\alpha,K\setminus\{j\})C_{j,k} & \text{if } j,k\in K \\
		C_{j,k}G_Z(\alpha,K)=G_Z(\alpha,K\cup\{j\})C_{j,k} & \text{if } j\notin K,\text{ }k\in K \\
		C_{j,k}G_Z(\alpha,K)=G_Z(\alpha,K)C_{j,k} & \text{if } k\notin K.
	\end{cases}.\label{eqCommutationCX}
\end{equation}
and similarly, we have,
\begin{equation}
	\begin{cases}
		C_{k,j}G_X(\alpha,K)=G_X(\alpha,K\setminus\{j\})C_{k,j} & \text{if } j,k\in K \\
		C_{k,j}G_X(\alpha,K)=G_X(\alpha,K\cup\{j\})C_{k,j} & \text{if } j\notin K,\text{ }k\in K \\
		C_{k,j}G_X(\alpha,K)=G_X(\alpha,K)C_{k,j} & \text{if } k\notin K.
	\end{cases}.\label{eqCommutationCX2}
\end{equation}

\begin{figure}[h]
	\centering
	\begin{subfigure}[a]{1.0\textwidth}
		\includegraphics[width=\textwidth]{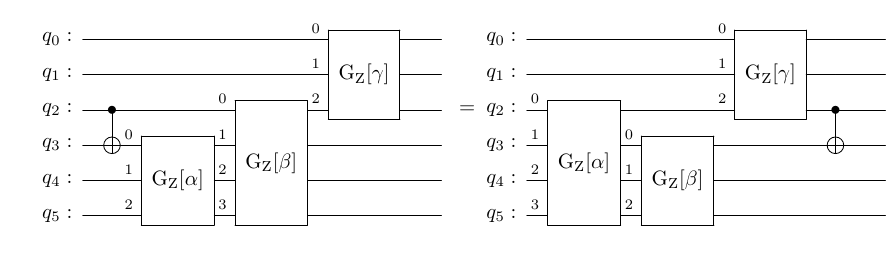}
	\end{subfigure}
	\begin{subfigure}[b]{1.0\textwidth}
		\includegraphics[width=\textwidth]{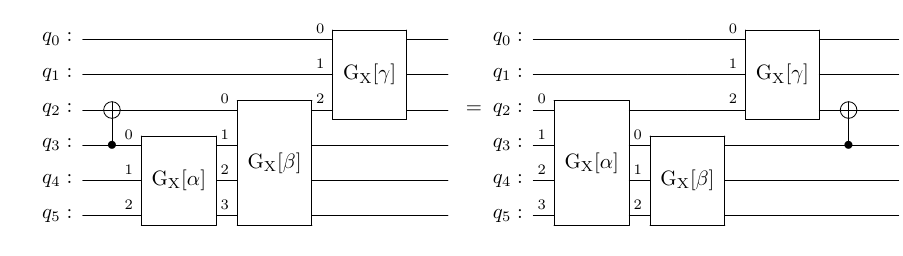}
	\end{subfigure}
	\caption{Top: Commutation rules examples for $Z$ phase gadgets. Bottom: Commutation rules examples for $X$ phase gadgets. \label{fig:CommRules}}
\end{figure}

Figure~\ref{fig:CommRules} (top) shows these relations, in which the control qubit is added (removed) from the PG in case the target qubit is included in the PG and the control qubit is not (is as well). When the target qubit is not included in the PG the operations commute. Similar relations are known for $X$-PGs, shown in Fig.~\ref{fig:CommRules} (bottom) as well, with the roles of $j$ and $k$ in the conditions of Eq.~\eqref{eqCommutationCX} exchanged. These relatively simple CNOT commutation rules are a special case of commutation of PGs with general Clifford operations \cite{cowtan2020phase}.

As all CNOTs on arbitrary qubit pairs and arbitrary single-qubit rotations form a universal gate set, and since single-qubit gates can be written in terms of single-qubit $Z$ and $X$ PGs, then the CNOT commutation rules are sufficient to realize the three-layer structure described above. Namely, a circuit can be described with this gate set and all CNOT gates can be commuted to the beginning or end of the gate, one-by-one, and in this process modify PGs, as is detailed below. 

We remark that an additional tool used to reduce the number of PGs is a commutation relation of PGs, namely that $\left[G_X(\alpha,J),G_Z(\beta,K)\right]=0$ if $|J\cap K|$ is even (see the SM \cite{SM}).

\section{Layering and $SU(4)$ decomposition}\label{sec:SU4}

As we show below, the number of single qubit gates affect the depth of the compiled circuit, thus it is useful to absorb such redundant gates. A basic routine to do so, used throughout our compilation, is composition to layers and their subsequent decomposition. For the composition, we generate `vertical' layers of gates such that all of the gates within each layer commute. This is done by passing through all the gate-instruction of a given input circuit and performing the following per gate: assuming already existing $\ell$ layers, we check whether the gate commutes with all gates in layer $\ell$, then $\ell-1$, then $\ell-2,...$ etc. and we place the gate at layer $\ell^\ast+1$ (or create it) such that $\ell^\ast$ is the largest layer index with which the gate doesn't commute. After this process concludes we generate from the layers $SU(4)$ blocks by pairing the qubits according to the gates appearing in the layer.

After all $SU(4)$ operators are formed we decompose them back to single-qubit and two-qubit gates. There are well-known methods to do so, such as the Cartan decomposition \cite{navin2001time,tucci2005introduction,zhang2003geometric,blaauboer2008analytical}, however this generically results in 9 layers of single qubit rotations, which can be asymptotically reduced to 8 when considering consecutive decompositions. Instead, we use the left-handed (LH) decomposition, introduced in Ref. \cite{nemirovsky2025efficient}, in which $SU(4)$ operators are decomposed, using up to three $Z\otimes Z$ rotations, with one of them appearing as the first gate of the block. This decomposition incurs only 7 single-qubit layers, which is similarly asymptotically reduces to 6 layers, reducing the resulting compiled circuit depth. 

An additional optimization is performed to reduce the nuclear norm of the decomposition. This is done by adding identical CNOT pairs to the end of the $SU(4)$ block, which amounts to an identity operator. However only one of these CNOT gates is included in the LH decomposition and the other is left unchanged. There are 6 different combinations of CNOT operators ($I$, $C_{n,m}$, $C_{m,n}$, $C_{n,m}C_{m,n}$, $C_{m,n}C_{n,m}$, SWAP), which result in different total entanglement phases of the three $Z\otimes Z$ rotations (in absolute value), which eventually contribute to the total nuclear norm of the circuit, and the remaining CNOTs can be commuted to the circuit edges, detailed below. Thus, we check all of these combinations, per block, and choose the one with the least total entanglement phase.

\section{Compilation primitive}\label{sec:CompPrim}
We describe our main compilation primitive, $\text{PG}_\text{L}\left(U\right)$, which is then used as an algorithmic block in optimization passes (here `L' denotes left-handedness). Assuming an input circuit, $U$, we first perform layering and decomposition as detailed above. Next, we `pull' all of the CNOT gates to the beginning of the circuit. this is done by commuting the first CNOT gate through all the PGs appearing before it, using the commutation rules in Eqs.~\eqref{eqCommutationCX} and~\eqref{eqCommutationCX2} above. Similarly, the second CNOT gate is commuted, etc. At this point the circuit is decomposed to a layer of CNOT gates followed by a layer of PGs, where the number of PGs is bounded by the number of single qubit gates appearing in the layer decomposition.

Next, some trivial optimization steps are performed on the PGs in order to reduce their number and the nuclear-norm required to implement them. Namely, PGs can be commuted according to the identity above, consecutive PGs over the same Pauli axis and qubit sets can be merged, rotation angles can be kept between $-\pi/4$ and $\pi/4$ by adding single qubit gates, and trivial PGs can be removed.

We perform an additional non-trivial global optimization over the nuclear-norm, by adding CNOT gate pairs to the beginning of the circuit and commuting one of them to its end. Specifically, for a PG circuit, $U_\text{PG}$, we consider the representation $U_\text{PG}=C_{n,m}U_{n,m}^\ast C_{n,m}$ with $U_{n,m}^\ast=C_{n,m}U_\text{PG}C_{n,m}$. Clearly, $U_{n,m}^\ast$ is as well a PG circuit, however the sets of $J$s are modified, and may contain less qubits, thus incurring a smaller nuclear norm. Luckily, the commutation and implementation rules detailed above make the computation of the modified nuclear norm simple. 

We preform this optimization in steps, such that in each step we form a $N\times$N matrix with the $(n,m)$ entry corresponding to the nuclear-norm of $U_{n,m}^\ast$ and implement a set of commuting CNOT gates that act to reduce the nuclear norm. This is repeated until no more reduction is possible. By representing the set of CNOT operators in their symplectic form \cite{aaronson2004improved,nemirovsky2025reduced}, then these optimization steps can be represented as matrix multiplications, making them amenable to implementation on graphic processing units (GPUs).

We note that the optimal choice of CNOT gates forms an instance of the maximum weight matching problem, which has well-known and efficient heuristic solutions~\cite{edmonds1965paths}.

\section{Global optimization passes}\label{sec:CompGlob}
The compilation primitive above compiles circuits into a layer of CNOT gates followed by a layer of PGs, $\text{PG}_\text{L}\left(U_\text{input}\right)=\left(U_\text{PG},U_\text{C}\right)$, such that $U_\text{input}=U_\text{PG}U_\text{C}$. Interestingly, this factorization is `unstable' in the sense that $\text{PG}_\text{L}\left(U_\text{PG}\right)\neq\left(I,U_\text{PG}\right)$, meaning that optimization passes can be performed in iterations. Furthermore, we analogously define a right-handed primitive, $\text{PG}_\text{R}\left(U_\text{input}\right)=\left(U_\text{C},U_\text{PG}\right)$, such that $U_\text{input}=U_\text{C}U_\text{PG}$, which makes use of the same techniques detailed above. In fact, one can implement this primitive with the left-handed primitive by operating on $U_\text{input}^\dagger$ and then conjugate the resulting $U_\text{PG}$.

These tools imply optimization passes that can be iteratively performed. We initialize the iteration by setting $U_\text{PG}^{\left(0\right)}=\text{argmin}\left\{\left|U_\text{PG,L}\right|,\left|U_\text{PG,R}\right|\right\}$, with $U_\text{PG,L}$ ($U_\text{PG,R}$) the phase gadget layers given by the left-handed (right-handed) compilation primitive operating on the input circuit, and the norm represents any desirable performance metric. Typically, the norm of choice is the $\umq$ gate count and total nuclear-norm, yet other choices exist, e.g. minimization of non-Clifford operations.

Next we perform iterations,
\begin{equation}
	U_\text{PG}^{\left(n+1\right)}=\text{argmin}\left\{\left|U_\text{PG,L}^{\left(n+1\right)}\right|,\left|U_\text{PG,R}^{\left(n+1\right)}\right|\right\}.\label{eqIterations}
\end{equation}
The iteration terminates when $\left|U_\text{PG}^{\left(n+1\right)}\right|\geq\left|U_\text{PG}^{\left(n\right)}\right|$. During these iterations CNOT layers are accumulated before and after the PG layer, and are trivially merged with existing CNOT layers. Lastly the compiler outputs $U_\text{PG}^{\left(n\right)}$ and the corresponding CNOT layers. An example of an input circuits and its resulting compilation is provided in SM \cite{SM}.

Notably, the compilation method's complexity has a polynomial scaling with circuit depth, meaning that typically circuits can be compiled as a whole. Nevertheless, for extremely deep circuits it might be useful to `cut' the circuit to a few blocks and optimize them independently. In this cases intermediate CNOT (or even Clifford) layers will be generated between the blocks, which cannot be processed with classical compute. Thus, these layers will be implemented with the quantum hardware, however with a small overhead of only few instances of $\umq$, as shown in Ref. \cite{nemirovsky2025reduced}.
 
\section{Noise models}\label{sec:Noise}
To benchmark the efficacy of our compilation method we consider non-ideal hardware realizations of the input and compiled circuits, and compare their performance. We consider relevant noise models for the two-qubit and multiqubit entanglement gates: qubit depolarization due to photon scattering and qubit phase-noise. These noise channels are the leading sources of error in high-fidelity realization of trapped-ions entanglement gates \cite{ballance2016high}. We neglect single-qubit gate errors as these are generally subdominant \cite{smith2025single}.

Circuit errors are introduced via noise injection, i.e. we first determine the error probability per noise channel and per entangling gate, then, we sample many noisy realizations according to these probabilities. Specifically, the dephasing channel simply introduces a fixed error probability per entanglement gate (two-qubit and multiqubit alike), such that independently for each qubit a $Z$-operator is introduced before the operation of the gate with probability $p_\text{dephase}$. Here we fix, $p_\text{dephase}=10^{-3}$.

Depolarization is more involved as it typically depends on the gate drive power, e.g. in the case of unwanted photon-scattering. Thus we first estimate the drive power using the nuclear norm of $\varphi$, as described above, and use it to `gauge' the error probability. To be concise, we fix the error probability of a fully entangling $Z\otimes Z$ two-qubit gate, $p_\text{depol,TQ}$, and derive the depolarization error probability of any other gate by considering the nuclear norm ratio and the qubit participation. Such a method has also been employed in Ref. \cite{nemirovsky2025efficient}. We then introduce independently and uniformly for each qubit a $X$, $Y$ or $Z$ operation with probability $p_\text{depol,MQ}$ before the operation of the gate. Here we fix the fully entangling two-qubit gate error to $p_\text{depol,TQ}=10^{-3}$, which leads, e.g., for the all-to-one gates used in Eq.~\eqref{eqPGdecomp} to a multiqubit error probability of $p_\text{depol,MQ}\approx0.00528$ in a $N=30$ register (and an asymptotic $\sim\sqrt{N}$ scaling in general).

With these settings qubit depolarization is sensitive to gate count and qubit depolarization is sensitive to total nuclear norm. Here we choose to set these two error channels on-par, which is realized by setting $p_\text{dephase}=p_\text{depol,TQ}$ \cite{ballance2016high}. 

A simple estimation of performance is formed by setting the realization fidelity to the probability of sampling an ideal, noise-less, circuit. This over-estimates the effects of noise as it assumes that any non-ideal instance has a vanishing fidelity. A more accurate, yet compute-intensive, estimate is formed by sampling a large ensemble of noisy realizations of the circuits and performing state-vector simulation shots on these ensembles.

With this, we reconstruct the bit-string probability distribution function (up to shot noise), and set the circuit fidelity according to the total variation distance, i.e. 
\begin{equation}
	F_\text{TVD}=1-\frac{1}{2}\sum_x \left|p_\text{ideal}(x)-p_\text{sim}(x)\right|,\label{eqSim}
\end{equation}
with $p_\text{ideal}(x)$ ($p_\text{sim}(x)$) the noise-less (reconstructed) probability of bit string $x$. In our benchmarks below we show that both methods yield consistent results.

\section{Numerical benchmarks}\label{sec:Bench}

\begin{figure}[h]
	\centering
	\includegraphics[width=1\textwidth]{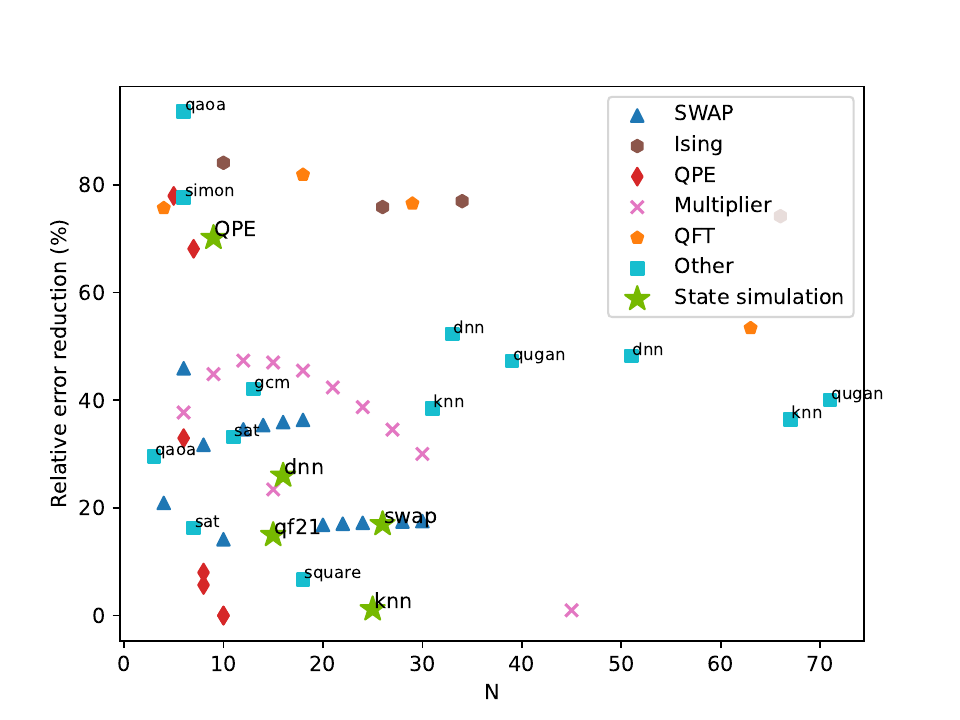} % no extension if PDF/PNG
	\caption{Relative error reduction factor, according to Eq.~\eqref{eqRelError} (vertical). We consider various circuits for varying number of qubits, $N$ (horizontal), grouped by circuit type: Swap test (blue triangles), Ising model simulation (brown hexagons), Quantum phase estimation algorithm (red diamonds), Number multiplication (pink crosses), quantum Fourier transform (orange pentagons) and others (cyan squares, see caption). A generic relative error reduction of $\sim40\%$ is observed. A subset of circuits is tested with noise-injection and state simulation (green stars) showing an agreement with the other circuits.}
	\label{fig:rel_error}
\end{figure}

We benchmark our compilation by testing it on a large set of quantum circuits, some of which obtained from the QASMBench repository \cite{li2023qasmbench}. As shown below, some of the circuits are part of a series, synthesized for various qubit register sizes: Swap test (`Swap'), Ising model simulation ('Ising'), Quantum phase estimation algorithm (`QPE'), Number multiplication (`Multiplier') and quantum Fourier transform (`QFT'). Other circuits are specific to a fixed register size: Square root (`square'), quantum neural networks (`qugan' and `dnn'), quantum K-nearest neighbor (`knn'), Quantum approximate optimization algorithm (`qaoa'), factorization (`qf21'), Simon’s algorithm (`simon'), Boolean satisfiability problem (`sat'), generator coordinate method (`gcm') and quantum error correction (`qec9xz'). 

For each input circuit from the set we consider several relevant metrics for the optimization, and present them for each circuit as well as mean values. Since the means depend strongly on the chosen circuits and lack strict statistical rigor, they are presented solely as a qualitative indicator of relative performance among configurations. Here we present our results graphically, while the full numerical data are provided in the SM \cite{SM}.

We first consider the most important metric - the relative error reduction factor, given by,
\begin{equation}
	\epsilon_\text{relative}=\frac{F_\text{comp}-F_\text{inp}}{1-F_\text{inp}},\label{eqRelError}
\end{equation}
where $F_\text{comp}$ and $F_\text{inp}$ are the circuit fidelities of the compiled and input circuits, respectively. The fidelities are calculated either by a success probability estimation or by state-simulation, as detailed above.

Figure \ref{fig:rel_error} presents relative error reduction factor for the benchmark circuit set. For most circuits the compilation reduces the error, exhibited by a positive error reduction factor, $\epsilon_\text{relative}$, with an overall average relative error reduction of $\sim 40\%$. For a small number of instances no reduction is observed, implying PG compilation did not reduce errors over these instances. For a limited set of circuits we use the more accurate evaluation method, making use of noise-injection and state-simulation over at least 10,000 different noise samples, each (green stars). These results are as well consistent with entire data set.

\begin{figure}[h]
	\centering
	\includegraphics[width=1\textwidth]{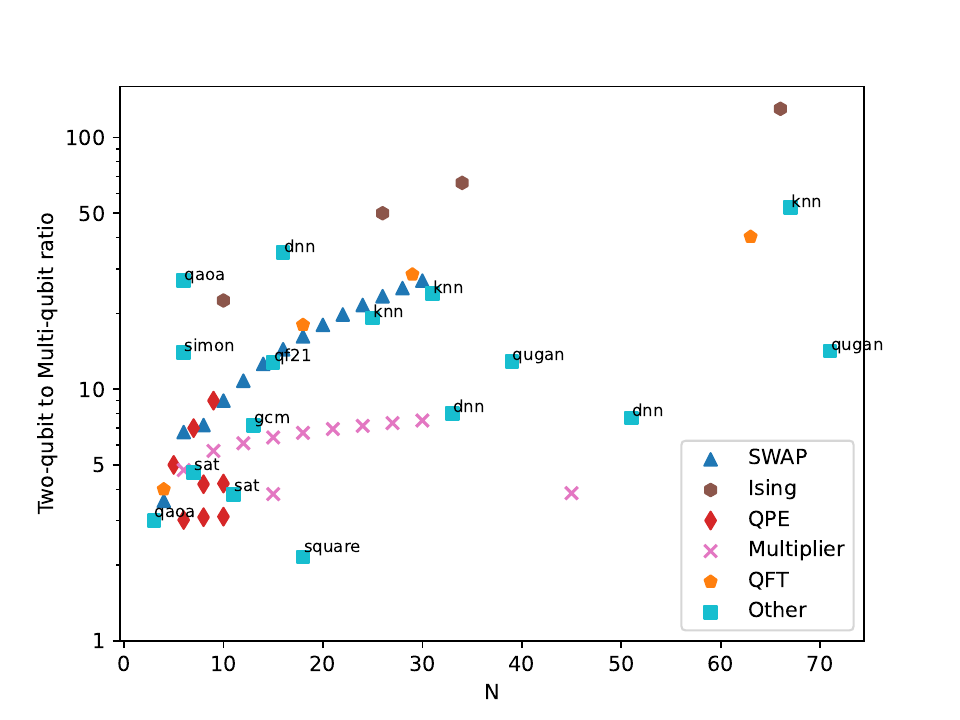} % no extension if PDF/PNG
	\caption{Two-qubit gate count to compiled multiqubit gate count ratio due to PG compilation (vertical, log-scale). We consider various circuits for varying number of qubits, $N$ (horizontal), grouped by circuit type as in Fig. \ref{fig:rel_error}. A generic gate count reduction of $\sim15$ is observed (reaching up to $\sim100$ in some cases). For Swap test (blue triangles), Ising model simulation (brown hexagons) and quantum Fourier transform (orange pentagons) performance clearly improves as $N$ increases.}
	\label{fig:two_to_multi}
\end{figure}

We are interested in how the error reduction comes about. For this, we first consider depth reduction. Figure \ref{fig:two_to_multi} shows the ratio of the number of two-qubit gates of input circuits to the number of multiqubit gates of the compiled circuits. The two-qubit gate count is obtained by straightforwardly converting all entanglement gates in the input circuit to $Z\otimes Z$-type entanglement operations, while the multiqubit gate count is obtained from the middle PG layer of our compilation method. We observer that the number of gates has been reduced in all circuit (for some even by a factor of $\sim100$), with an average gate reduction factor of $\sim 15$.

\begin{figure}[h]
	\centering
	\includegraphics[width=1\textwidth]{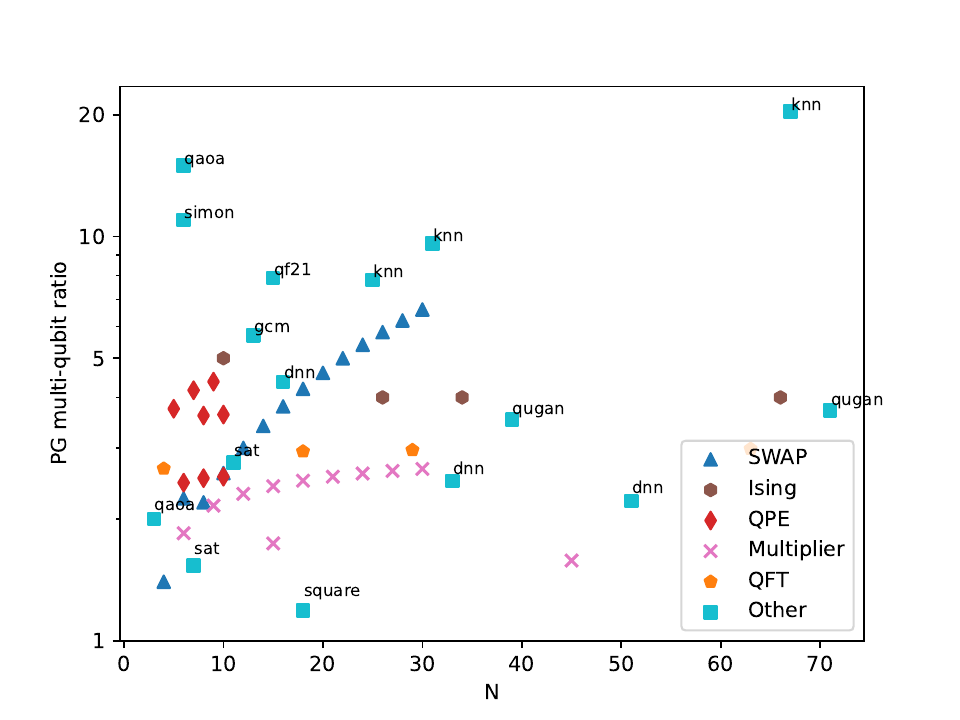} % no extension if PDF/PNG
	\caption{Multi-qubit gate count ratio of a straightforward use of multi-qubti gates to the PG compiled circuits gate count (vertical, log-scale) for varying number of qubits, $N$ (horizontal), grouped by circuit type as in Fig. \ref{fig:rel_error}. An overall average reduction by a factor of $\sim4$ is attributed to PG compilation.}
	\label{fig:multi_to_PG}
\end{figure}

It is revealing to separate the roles of generically using $\umq$ and additionally using PG compilation. For this we consider the input circuits (after conversion to $Z\otimes Z$-type gates) and merge parallel entanglement operations using $\umq$, i.e. without performing non-trivial gate commutations. We compare the multiqubit gate count of this straightforward approach to the PG compilation. Figure \ref{fig:multi_to_PG} shows the results of this analysis, with an average depth reduction factor of $\sim4$. Accordingly, this factor results from the performance of PG compilation.

\begin{figure}[h]
	\centering
	\includegraphics[width=1\textwidth]{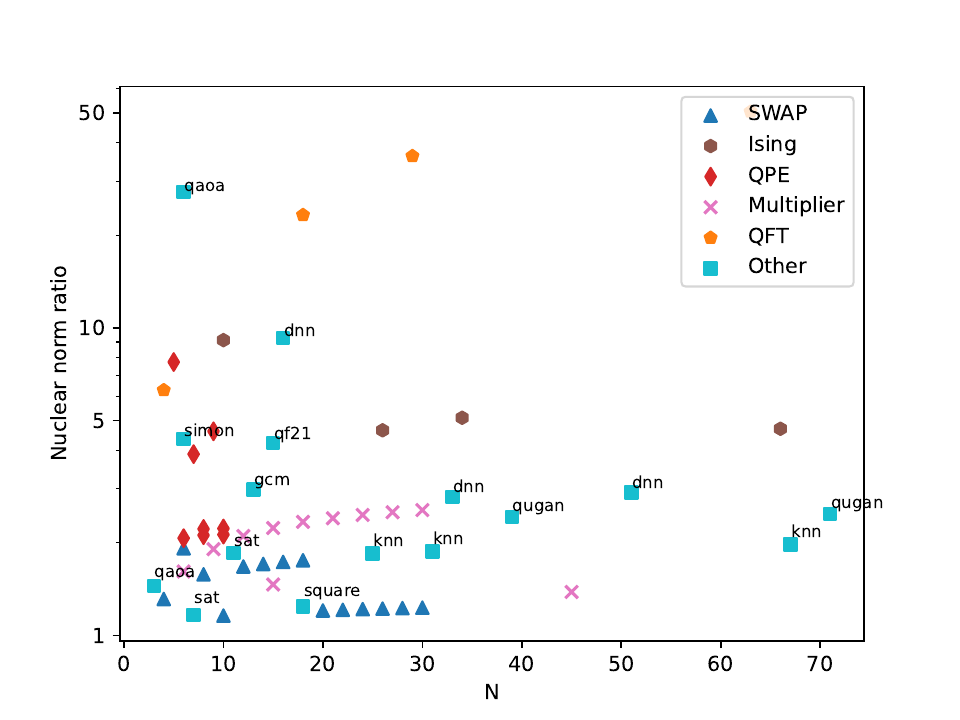} % no extension if PDF/PNG
	\caption{Total nuclear norm reduction factor, comparing the ratio of the total nuclear of the input circuit to that of the PG compiled circuit (vertical), for varying number of qubits, $N$ (hotizontal), grouped by circuit type as in Fig. \ref{fig:rel_error}. We observe an overall average reduction in gate drive power by a factor of $\sim4$.}
	\label{fig:nuc_ratio}
\end{figure}

Next we consider the reduction in nuclear norm, which is related to the multiqubit gate drive power and subsequent error. For the converted-input gate we compute the total nuclear norm of all of the $Z\otimes Z$ gates, while for the PG compiled circuits we consider the total nuclear norm of all instances of $\umq$, after performing the optimization outlined above. Figure \ref{fig:nuc_ratio} shows the results of this analysis, with an average nuclear norm reduction factor of $\sim4$. That is, the depth reduction and accompanying improved execution time do not incur additional gate drive power, rather it is typically reduced. In some cases, e.g. `QAOA', both depth and nuclear norm are significantly reduced.

\section{Conclusions}

In conclusion, we have presented a method that leverages a phase-gadget formalism and programmable multiqubit entanglement gates, in order to provide generic circuit depth reduction by a factor of $\sim15$ and drive-power reduction by a factor of $\sim4$. We performed benchmarks of our method on a large set of circuits, showing that these ultimately reduce the relative implementation error of quantum circuits by $\sim40\%$. As the implementation of phase-gadgets does not saturate the number of programmable pairwise couplings, we expect that long-range all-to-all interactions will enable further speed-ups \cite{yin2025fastquantumcomputationalltoall}.

\begin{acknowledgments}
We thank Noa Pariente and Bar Segal for their support in providing the HPC infrastructure. 
\end{acknowledgments}

\bibliography{references}

\cleardoublepage
\onecolumngrid
{
	\begin{center}
	{\large \bfseries Supplemental material \par}
	\end{center}	
	\bigskip
	\setcounter{section}{0}
	{\section{Phase gadget relations}
In this section we provide comprehensive derivations 
and proofs for the Phase gadget rules presented in Eqs. (\ref{eqPGdecomp}), (\ref{eqPGmerge}) and (\ref{eqCommutationCX}).
We remark that Ref. \cite{cowtan2019phase} provides a good exposition for some of these relations.

\subsection{Proof of Eq. (\ref{eqPGdecomp}) - phase gadget decomposition}

Let $x_0, x_1,...x_n \in \{0,1\}$ be an arbitrary bit string and let $\ket{ x_0,x_1,...x_n }$ denote the common eigenstate of the mutually commuting Pauli operators $Z_0, Z_1,...,Z_n$, with $x_0, x_1,...x_n$ being the eigenvalues of these operators (resp.).
The $2^{n+1}$ ket states $\ket{ x_0,x_1,...x_n }$ are mutually orthogonal and could therefore be used to define any multi-qubit gate operating on qubits $0,1,...,n$.
Recalling that CNOT gates act as XOR operations on these ket states, as follows,
\begin{equation}
	C_{Z_j\land X_0} \ket{ x_0,x_1,...,x_n }=\ket{ x_0\oplus x_j,x_1,...,x_n },
\end{equation}
with the symbol "$\oplus$" denoting a binary XOR operation, we obtain,
\begin{equation}
	\left(\prod^n_{j=1}C_{Z_j\land X_{0}}\right)\ket{ x_0,x_1,...,x_n }=\ket{ x_0\oplus x_1\oplus ...\oplus x_n,x_1,...,x_n },
\end{equation}
\begin{equation}
 Z_0^\alpha\left(\prod^n_{j=1}C_{Z_j\land X_{0}}\right)\ket{ x_0,x_1,...,x_n }=e^{i\alpha\pi\left(x_0\oplus x_1\oplus ...\oplus x_n\right)}\ket{ x_0\oplus x_1\oplus ...\oplus x_n,x_1,...,x_n }
\end{equation}
and
\begin{align}
	\left(\prod^n_{j=1}C_{Z_j\land X_{0}}\right)Z_0^\alpha\left(\prod^n_{j=1}C_{Z_j\land X_{0}}\right)\ket{ x_0,x_1,...,x_n }= & e^{i\alpha\pi\left(x_0\oplus x_1\oplus ...\oplus x_n\right)}\ket{ x_0,x_1,...,x_n }\equiv \nonumber \\
	& \equiv P_Z(\alpha,\{x_0,x_1,...,x_n\})\ket{ x_0,x_1,...,x_n }, \label{eq:decomp_of_PG}
\end{align}
where the last equality in the right hand side of Eq. (\ref{eq:decomp_of_PG}) is actually the definition of a $Z$-phase gadget gate.
Finally, due to the orthonormality of the ket states $\ket{ x_0,x_1,...x_n }$ we conclude that,
\begin{equation}
	\left(\prod^n_{j=1}C_{Z_j\land X_{0}}\right)Z_0^\alpha\left(\prod^n_{j=1}C_{Z_j\land X_{0}}\right)=P_Z(\alpha,\{x_0,x_1,...,x_n\}).
	\label{eq:PG_parity_rule}
\end{equation}

Similarly, conjugating Eq. (\ref{eq:PG_parity_rule}) with $e^{-i\frac{\pi}{4}Z_0}$ - i.e., a $\frac{\pi}{2}$ rotation of qubit 0, we obtain,
\begin{align}
	P_Z(\alpha,\{x_0,x_1,...,x_n\}) &=  e^{-i\frac{\pi}{4}Z_0} P_Z(\alpha,\{x_0,x_1,...,x_n\}) e^{i\frac{\pi}{4}Z_0} = \nonumber \\
	 & =e^{-i\frac{\pi}{4}Z_0}\left(\prod^n_{j=1}C_{Z_j\land X_{0}}\right)Z_0^\alpha\left(\prod^n_{j=1}C_{Z_j\land X_{0}}\right)e^{i\frac{\pi}{4}Z_0}
	=\left(\prod^n_{j=1}C_{Z_j\land Y_{0}}\right)Z_0^\alpha\left(\prod^n_{j=1}C_{Z_j\land Y_{0}}\right)
\end{align}

Finally, permuting the axes in a cyclic manner according to $ Z \rightarrow X \rightarrow Y \rightarrow Z $ reveals all of the various versions of the rule stated in Eq. (\ref{eqPGdecomp}).

\subsection{Proof of Eq. (\ref{eqPGmerge}) - phase gadget merge}
Let $x_0, x_1,...x_n \in \{0,1\}$ be an arbitrary bit string and let $\ket{ x_0,x_1,...x_n }$ denote the common eigenstate of the mutually commuting Pauli operators $Z_0, Z_1,...,Z_n$, with $x_0, x_1,...x_n$ being the eigenvalues of these operators (resp.).
Denote with $a=0$ the auxiliary qubit index and the rest of the qubits with the set $ J=\{1,2,3,...,n\} $ we obtain,
\begin{align}
	\prod_{j\in J}C_{X_j\land Z_a} \prod_{k\in J}C_{Y_k\land Z_a} &   \ket{ x_a,x_1,...,x_n } = \nonumber \\
	& =\prod_{j\in J}C_{X_j\land Z_a}\left(\prod_{k\in J}  e^{i\frac{\pi}{2}x_k \oplus x_a} e^{-i\frac{\pi}{2}x_k}\right) \ket{ x_a,x_1\oplus x_0,...,x_n\oplus x_0 } =\nonumber \\
	& = \prod_{k\in J}  e^{i\frac{\pi}{2}x_k \oplus x_a} e^{-i\frac{\pi}{2}x_k} \ket{ x_a,x_1,...,x_n }  =\nonumber \\
	& = \prod_{k\in J} Z_k^{-1/2} G_Z(\frac{\pi}{2},\{a,k\})\ket{ x_a,x_1,...,x_n } .
		\label{eq:merge_proof}
\end{align}
Where in the above relations, we have used the rule,
\begin{align}
	C_{Y_k\land Z_a} \ket{ x_a,x_1,...,x_n } & = e^{-i\frac{\pi}{4}Z_k}C_{X_k\land Z_a}e^{i\frac{\pi}{4}Z_k} \ket{ x_a,x_1,...,x_n } = \nonumber \\
	 & = e^{i\frac{\pi}{2}x_k \oplus x_a} e^{-i\frac{\pi}{2}x_k} \ket{ x_a,x_1,...,x_k \oplus x_a,...,x_n } .
\end{align}

As Eq. (\ref{eq:merge_proof}) is true for all $2^{n+1}$ elements in the orthogonal set $\{ \ket{ x_a,x_1,...,x_n } | x_a,x_1,...,x_n\in\{0,1\}\}$, we conclude that,
\begin{equation}
	\prod_{j\in J}C_{X_j\land Z_a} \prod_{k\in J}C_{Y_k\land Z_a} =  \prod_{k\in J} Z_k^{-1/2} G_Z\left(\frac{\pi}{2},\{a,k\}\right) .
\end{equation}
Finally, performing a cyclic permutation of the axes $X\rightarrow Y\rightarrow Z\rightarrow Z\rightarrow X$, we obtain,
\begin{equation}
	\prod_{j\in J}C_{Z_j\land Y_a} \prod_{k\in J}C_{X_k\land Y_a} =  \prod_{k\in J} Y_k^{-1/2} G_Y\left(\frac{\pi}{2},\{a,k\}\right)
\end{equation}
and applying complex conjugate on both sides of the last equation yields the other version of the merge rule,
\begin{equation}
	\prod_{j\in J}C_{X_j\land Y_a} \prod_{k\in J}C_{Z_k\land Y_a} =  \prod_{k\in J} Y_k^{1/2} G_Y\left(-\frac{\pi}{2},\{a,k\}\right) .
\end{equation}
This completes the proof of the merge rule stated in Eq. (\ref{eqPGmerge}).

\subsection{Proof of Eqs. (\ref{eqCommutationCX}-\ref{eqCommutationCX2}) - phase gadget and CNOT commutation rules}

To prove the first and second rules stated in Eqs. (\ref{eqCommutationCX}-\ref{eqCommutationCX2}) for $j,k\in K$ we use Eq. (\ref{eqPGdecomp}) to decompose a phase gadget $P_Z(\alpha,K)$ as follows,
\begin{equation}
	P_Z(\alpha,K) = \left(\prod_{\ell\in K\backslash \{k\}}C_{Z_\ell\land X_k}\right)Z_k^\alpha\left(\prod_{\ell\in K\backslash \{k\}}C_{Z_\ell\land X_k}\right),
	\label{eq:temp}
\end{equation}
Conjugating the left and right hand-side of Eq. (\ref{eq:temp}) with the control gate $C_{Z_j \land X_k}$ where $j\in K$ yields,
\begin{equation}
	C_{Z_j \land X_k} P_Z(\alpha,K) C_{Z_j \land X_k} = 
	\left(\prod_{\ell\in K \backslash \{j,k\}} C_{Z_\ell\land X_k}\right)Z_k^\alpha\left(\prod_{\ell\in K \backslash \{j,k\} }C_{Z_\ell\land X_k}\right)= P_Z(\alpha,K\backslash\{j\}).
	\label{eq:conjugateMe}
\end{equation}
This concludes the proof of the first rule stated in Eq. (\ref{eqCommutationCX}).

Conjugating the left most and right most sides of Eq. (\ref{eq:conjugateMe}) with the control gate $C_{Z_j \land X_k}$ yields,
\begin{equation}
P_Z(\alpha,K)  = 
		C_{Z_j \land X_k} P_Z(\alpha,K\backslash\{j\})C_{Z_j \land X_k},
\end{equation}
proving the second rule stated in Eq. (\ref{eqCommutationCX}).

Finally, to prove the third rule of Eq. (\ref{eqCommutationCX}), i.e., the case in which $k \notin K$, we choose a $k' \in K\backslash \{j\}$ and note that,
\begin{align}
	C_{Z_j\land X_k} P_Z(\alpha,K) =C_{Z_j\land X_k}  & \left(\prod_{\ell\in K \backslash {k'}}C_{Z_\ell\land X_{k'}}\right)Z_{k'}^\alpha\left(\prod_{\ell\in K \backslash {k'}}C_{Z_\ell\land X_k}\right)= \nonumber\\ 
	& =\left(\prod_{\ell\in K \backslash {k'}}C_{Z_\ell\land X_{k'}}\right)Z_{k'}^\alpha\left(\prod_{\ell\in K \backslash {k'}}C_{Z_\ell\land X_k}\right)C_{Z_j\land X_k}= P_Z(\alpha,K)C_{Z_j\land X_k}.
\end{align}

\subsection{Proof that $[G_X(\alpha,J)$ , $G_Z(\beta,K)]=0$ if $ |J \cap K| $ is even - as stated at the end of section \ref{sec:Comm}}
Below we prove that phase gadgets $G_{P_1}(\alpha,J)$ and $G_{P_2}(\beta,K)$ commute, if $ J $ and $ K $,
 their sets of qubit indices share an even number of common indices - i.e., $|J\cap K|$ is even.
Note that this is true even if the Pauli operators $P_1$ and $P_2$ don't commute.
 
If $ P_1=P_2 $ the statement is trivial, so we'll prove first the case $P_1=X$ and $P_2=Z$ and remaining cases will follow with trivial axes rotations.

We start our proof by splitting the group of common qubit indices $ J \cap K $ into a pair of two disjoint set of indices $I_c$ and $I_t$,
having the same size, $|I_c|=|I_t|$ so that, $ J \cap K = I_c \cup I_t$ and $I_c \cap I_t=\emptyset$.
Furthermore, we enumerate the indices of these disjoint sets, so that $I_c=\{c_1, c_2, ... c_N\} $ and $I_t=\{t_1, t_2, ... t_N\}$ where $N=|I_c|=|I_t|$.
%as $|I_c|=|I_t|$ we identify pairs of indices $(c_\ell,t_\ell)$ where $c_\ell \in I_c$, $t_\ell\in I_t$ for $\ell = 1,2,...,|I_c|=|I_t|$.
Finally, note that exploiting the commutation rules of Eq. (\ref{eqCommutationCX}) we obtain,
\begin{align}
	G_X(\alpha,J)G_Z(\beta,K) & = \left(\prod^N_{\ell=1}  C_{Z_{c_\ell}\land X_{t_\ell}} \right)^2 G_X(\alpha,J)G_Z(\beta,K)\left(\prod^N_{\ell=1}   C_{Z_{c_\ell}\land X_{t_\ell}} \right)^2 = \\ = & \left(\prod^N_{\ell=1}  C_{Z_{c_\ell}\land X_{t_\ell}} \right) G_X(\alpha,J \backslash I_t )G_Z(\beta,K\backslash I_c)\left(\prod^N_{\ell=1}   C_{Z_{c_\ell}\land X_{t_\ell}} \right) = \\
	 = & \left(\prod^N_{\ell=1}  C_{Z_{c_\ell}\land X_{t_\ell}} \right)G_Z(\beta,K\backslash I_c) G_X(\alpha,J \backslash I_t )\left(\prod^N_{\ell=1}   C_{Z_{c_\ell}\land X_{t_\ell}} \right) = \\
	= & G_Z(\beta,K) G_X(\alpha,J  ).
\end{align}
\newpage
\section{Compilation example of a 6-qubit QAOA circuit from QASMBench \cite{li2023qasmbench}}
We consider the 6-qubit QAOA circuit, obtained from QASMBench \cite{li2023qasmbench}. Figure \ref{figE:PG_with_CNOTS} shows the original input circuit (a) and the compiled circuit (b). A clear reduction in depth, and specifically entanglement layers, is observed.

\begin{figure}[h]
	\centering
	\begin{subfigure}[a]{1.0\textwidth}
		\centering
		\includegraphics[width=\textwidth]{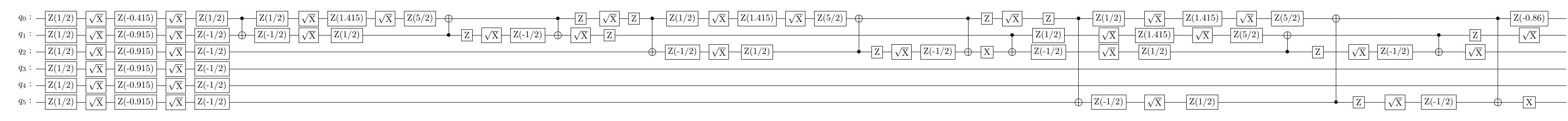}		$ \rightarrow $
		\includegraphics[width=\textwidth]{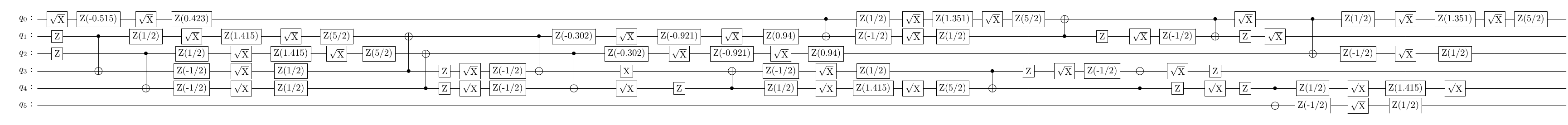}		$ \rightarrow $
		\includegraphics[width=\textwidth]{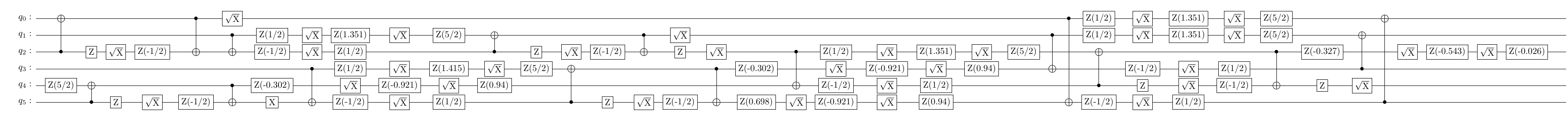}		$ \rightarrow $
		\includegraphics[width=\textwidth]{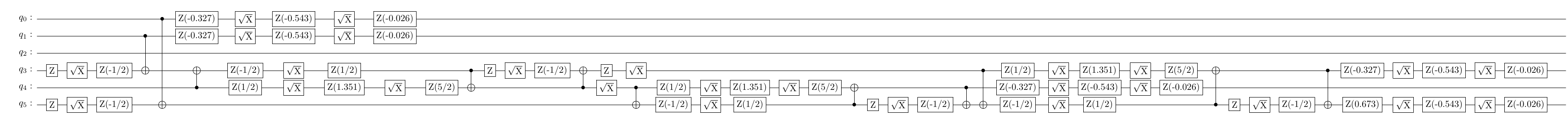}
		\caption{Input circuit.}
		\label{figE:sub1}
	\end{subfigure}
	\hfill
	
	\begin{subfigure}[b]{1.0\textwidth}
		\centering
		\includegraphics[width=\textwidth]{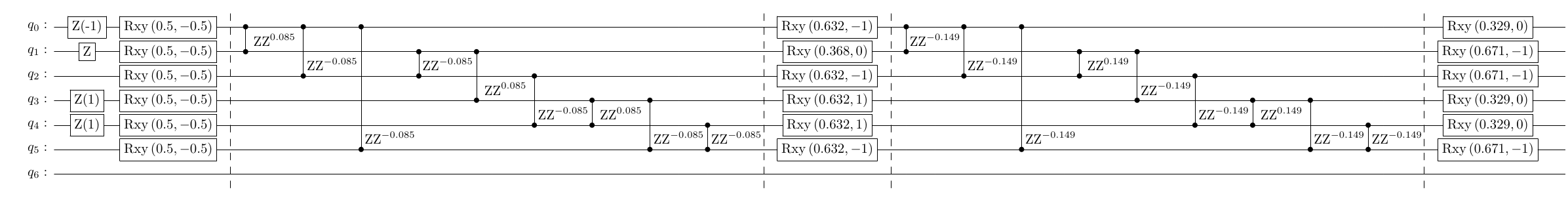}
		\caption{PG compiled circuit.}
		\label{figE:sub2}
	\end{subfigure}
	\caption{An example for phase gadget compilation of a six qubit QAOA circuit from QASMBench \cite{li2023qasmbench}.}
	\label{figE:PG_with_CNOTS}
\end{figure}

\newpage
\section{Detailed benchmark results}
We provide detailed numerical data on the benchmark circuit set. We compare the two-qubit and multiqubit gate count of the input and compiled circuits as well as their total nuclear norm and success probability, as detailed in the main text. We also provide relative improvement metrics. Additional general details regarding the circuit obtained from QASMBench appear in Ref.~\cite{li2023qasmbench}. In addition, we synthesize circuits for varying number of qubit registers, specifically a swap test circuit, number multiplication and quantum phase estimation for varying control and target qubits.

\begin{figure}[h]
	\centering
	\includegraphics[width=1\textwidth]{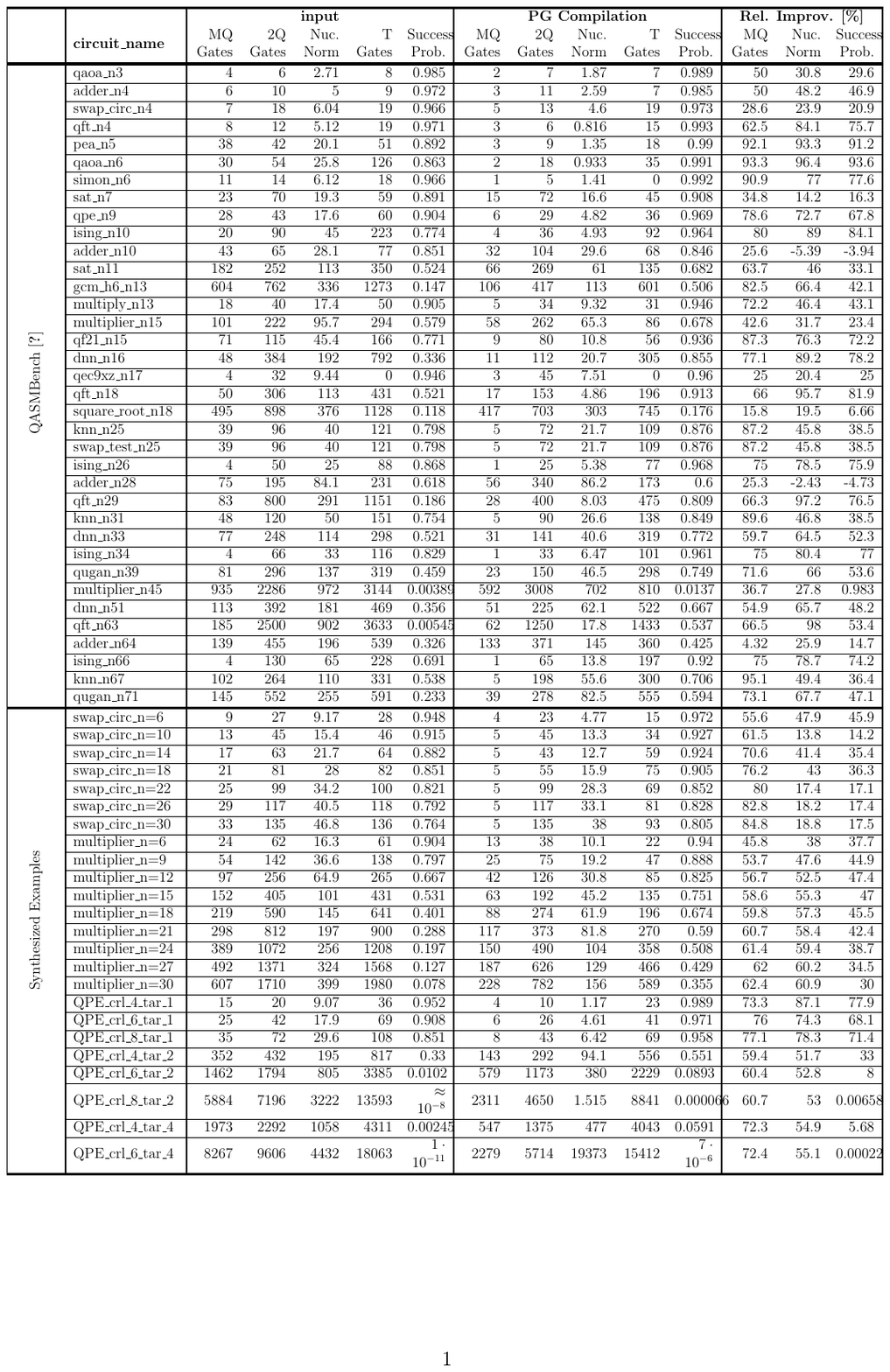} % no extension if PDF/PNG
\end{figure}

}
	\unskip
	}

\end{document}